\documentclass[aps,prb,amsmath,amssymb,twocolumn,superscriptaddress]{revtex4-1}
\usepackage{graphicx}
\usepackage{bm}
\usepackage[colorlinks,citecolor=blue,linkcolor=red]{hyperref}
\usepackage{color}

\begin{document}


\title{Zero-index and Topology in 1D Phononic Metamaterials with Negative Mass and Negative Coupling}

\author{Danmei Zhang}
 \affiliation{Center for Phononics and Thermal Energy Science, China-EU Joint Center for Nanophononics, Shanghai Key Laboratory of Special Artificial Microstructure Materials and Technology, School of Physics Sciences and Engineering, Tongji University, Shanghai 200092}

\author{Tianxiong Zhou}
 \affiliation{Center for Phononics and Thermal Energy Science, China-EU Joint Center for Nanophononics, Shanghai Key Laboratory of Special Artificial Microstructure Materials and Technology, School of Physics Sciences and Engineering, Tongji University, Shanghai 200092}

\author{Jie Ren}
\email{xonics@tongji.edu.cn}
\affiliation{Center for Phononics and Thermal Energy Science, China-EU Joint Center for Nanophononics, Shanghai Key Laboratory of Special Artificial Microstructure Materials and Technology, School of Physics Sciences and Engineering, Tongji University, Shanghai 200092}

\author{Baowen Li}
 \affiliation{Department of Mechanical Engineering, University of Colorado Boulder, Colorado 80309 USA}

\date{\today}

\begin{abstract}Phononic metamaterials have attracted extensive attention since they are flexibly adjustable to control the phonon transmission. In this work, we study a one-dimensional phononic metamaterial, made of mechanical resonant oscillators and chiral couplings.  
We show that by design, the oscillator mass and inter-oscillator coupling, although both are positive naturally, can be either single negative or double negative effectively within a certain frequency range. At the frequency where the effective mass and coupling are both infinite, a flat band emerges that will induce an extremely high density of states. At the critical point of band degeneracy, a Dirac-like point emerges where both effective mass and the inverse of effective coupling are simultaneously zero, so that zero index is realized for phonons. Moreover, the phononic topological phase transition is observed that the phononic band gap switches between single mass-negative and single coupling-negative regime. As a consequence, a topological interface state is identified, well explained by the theory.
\end{abstract}



\maketitle


\section{Introduction}
Metamaterials usually exhibit extraordinary properties that can not find in nature like the negative permittivity ($\varepsilon$), negative magnetic permeability ($\mu$) of electromagnetical metamaterials which can realize reversed Doppler effect, reversed Cerenkov radiation and negative refraction index~\cite{Veselago1968Veselago,pendry1996extremely,pendry1999magnetism,smith2000composite,Smith2000Negative,shelby2001experimental}. Along with the electromagnetical metamaterials for photonics, the phononic metamaterials have been greatly developed in recent years. In 2000, the phononic crystal with local resonance is proposed and the effective negative parameter is introduced~\cite{liu2000locally}. Then a variety of metamaterials comprising solid and liquid are proved to have double negative properties~\cite{li2004double-negative,Ding2007Metamaterial,Wu2011Elastic,bongard2010acoustic,Finocchio2014Seismic} which are reminiscent of Mie resonance. With the development of the theory, the negative modulus is demonstrated~\cite{fang2006ultrasonic} by the array of subwavelength Helmholtz resonators or side holes on a tube~\cite{Lee2008Acoustic} and the negative mass is realized by membrane-type acoustic tube~\cite{yang2008membrane-type} in low frequency range. After, the double negative parameters in the structure with array of periodic thin membranes and side holes~\cite{PhysRevLett.104.054301} are proposed. Then, non-periodic space-coiling structures~\cite{Liang2012Space,Xie2013Measurement} with double negativity are demonstrated to exhibit negative refractive index. In addition to the Mie resonance mechanism, there are also multiple scattering mechanisms leading to negative effective parameters~\cite{Torrent2008Anisotropic,Kaina2015Negative}.

The mass-spring structure with different coupling and spatial distribution can also have negative effective parameters\cite{Liu2016Elastic,lee2016origin,milton2007on,Yao2008Experimental,liu2011wave,Liu2011An,huang2011theoretical} because of Bragg scattering and resonant mechanism in different frequency regimes. In 2007, following the work of Ref.~\cite{liu2000locally}, Milton and Willis proposed a mass-in-mass system and showed that the dynamic effective mass of composite materials, defined in the framework of Newton's law of motion (contrary to the static gravitational mass), exhibits the existence of single or double negative properties~\cite{milton2007on}. In 2008, Yao $et$ $al.$ experimentally examined the model of mass-spring systems~\cite{Yao2008Experimental} with effective negative mass. Additionally in 2011, Liu $et$ $al.$ proposed an elastic model with double negative parameters by integrating a tri-chiral lattice with softly coated inclusions~\cite{liu2011wave} and an other double negative systems constructed by chiral mass-spring unit~\cite{Liu2011An}. Since then, a plenty of phononic metamaterials based on chiral and spiral structures are proposed~\cite{ZHU20142759,zhu2014negative,PhysRevLett.120.205501}.

On the other hand, there are many other properties of phononic metamaterials attracting increasing interest recent years, such as the zero refraction index (zero-index for short) and topological bands. In zero-index metamaterials, waves does not carry any spatial phase changes and the wave length is effectively infinite long~\cite{Engheta2013Pursuing} which can be used in wave surface modulation and bending waveguides. The zero-index material has been proved experimentally in the electromagnetic wave~\cite{Liu2008Experimental,Edwards2008Experimental} and the phononic counterparts have also been proved~\cite{Zheng2014Acoustic,Dubois2017Observation,Liu2012Dirac}. Meanwhile, over the past decades, the concept of ``topology'' has been attracting extensive research interests in phononics and phononic metamaterials~\cite{zhang2011the,PhysRevLett.105.225901,RevModPhys.84.1045,article22,PhysRevLett.119.255901}. Many interesting topological phenomena, such as topological interface and edge states~\cite{xiao2015geometric,Wang2015Topological}, have been observed in phononic metamatericals.

In this work, we study a 1D phononic metamaterials of mechanical resonant oscillators and chiral couplings. We analyze the phononic band structure by diagonalizing the dynamic matrix, calculate the effective mass and effective coupling by the equation of motion, and study the corresponding transmittance with transfer matrix method.
We show that by design, the oscillator mass and inter-oscillator coupling, although both are positive naturally, can be either single negative or double negative effectively within a certain frequency range. At the frequency where the effective mass and coupling are both infinite, a flat band emerges that will induce an extremely high density of states. Moreover, a Dirac-like point of phononic band emerges when both effective mass and the inverse of effective coupling are simultaneously zero, so that zero index is realized for phonons. Finally, we report the phononic topological phase transition that the phononic band gap switches between single mass-negative and single coupling-negative regime. Accordingly, a topological interface state is identified between two different single negative phononic materials.

\section{Model and Method}
\begin{figure}
\includegraphics[width=1\linewidth]{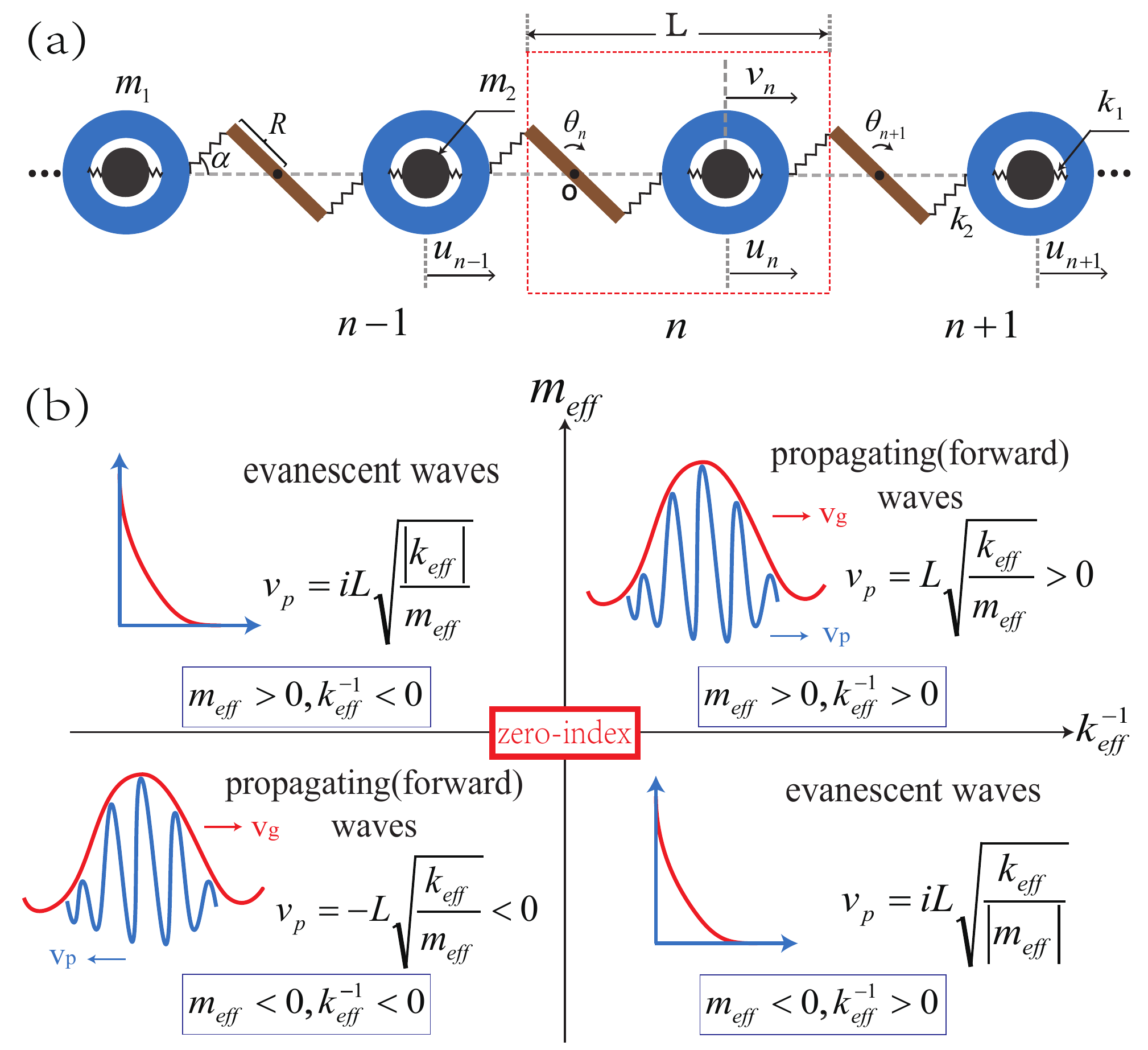}
\caption{\label{figure1}  (a) The one-dimension chiral lattice model with both negative mass and negative coupling.  (b) Classification of materials according to effective mass $m_{\rm{eff}}$ and inverse of effective spring coupling $k^{-1}_{\rm{eff}}$. $v_p$ is the phase velocity and $v_g$ is the group velocity.
}
\end{figure}

The system with resonant mass~\cite{milton2007on} and chiral spring coupling~\cite{Liu2011An,liu2011wave} is shown in Fig.~\ref{figure1}(a). The mass-in-mass unit takes the form of a rigid ball with mass $m_{1}$ and contains an internal mass $m_{2}$ that is connected to the outer ball by two internal spring $k_{1}$. In addition, we add a rigid leverage as the inter-unit coupling, whose moment of inertia and radius are $J$ and $R$. The middle of the leverage is fixed at the point O and  both ends are connected to the outer balls by the spring $k_{2}$. $\alpha$ is the equilibrium angle between $k_{2}$ and horizontal axis. The lattice constant of the system is $L$. For $m_{1}$ and $m_{2}$, only the movements in the horizontal direction $x$ are considered for simplicity.

Similar to photonic metamaterials, the phononic metamaterials can be generally classfied into four different categories according to the effective inverse coupling and mass $(k^{-1}_{\text{eff}}, m_{\text{eff}})$ shown in Fig.~\ref{figure1}(b). When they are double positive, the phase velocity is a positive number. The direction of phase velocity and group velocity are the same and the wave can propagate forward just like in the natural materials with normal refraction. But when one of $(k^{-1}_{\text{eff}}, m_{\text{eff}})$ is negative, we can see that the phase velocity is an imaginary number. According to the harmonic wave solution, the wave will present an evanescent wave of exponential decay. The third quadrant is when $(k^{-1}_{\text{eff}}, m_{\text{eff}})$ are both negative. In this case, the direction of phase velocity and group velocity is sign opposite and the wave can still propagate but with anomalous negative refraction. At the origin of the axes, the material possesses zero-index where the effective mass and the inverse of effective spring coupling are both zero, i.e. the refractive index equals to zero and the wave can propagate perfectly.

From the Newton's laws, the equations of motion for the $n$-th unit are given by
\begin{eqnarray}
\label{eq:a1}
m_{1}\frac{d^{2}u_{n}}{dt^{2}}&=&-2k_{1}(u_{n}-v_{n})
-k_{2}(u_{n}\cos\alpha-\theta_{n+1}R)\cos\alpha  \nonumber\\
&&-k_{2}(u_{n}\cos\alpha+\theta_{n}R)\cos\alpha,\\
\label{eq:a2}
m_{2}\frac{d^{2}v_{n}}{dt^{2}}&=&-2k_{1}(v_{n}-u_{n}),\\
\label{eq:a3}
J\frac{d^{2}\theta_{n}}{dt^{2}}&=&-k_{2}(2\theta_{n}R-u_{n-1}\cos\alpha+u_{n}\cos\alpha)R.
\end{eqnarray}
where $u_{n} (v_n)$ denotes the displacement of mass $1 (2)$ in the $n$-th cell, $\theta_{n}$ is the angle displacement of the leverage. 
Under harmonic excitation, as we all known, the harmonic wave solution of the 1D lattice chain is $(u_{n},v_{n},{\theta_{n}})=(\hat{u},\hat{v},\hat{\theta})e^{i(-nqL+{\omega}t)}$, with $q$ the quasi-momentum. So from Eqs. (\ref{eq:a2}) and (\ref{eq:a3}) we can have:
\begin{eqnarray}
\label{eq:a4}
v_{n}&=&\frac{2k_{1}u_{n}}{2k_{1}-m_{2}\omega^{2}}, \\
\label{eq:a5}
\theta_{n}&=&\frac{k_{2}R\cos\alpha}{J\omega^{2}-2k_{2}R^{2}}(u_{n}-u_{n-1}).
\end{eqnarray}
By substituting these equations into Eq.~(\ref{eq:a1}), we can reorganize the equation and obtain the following result:
\begin{equation}
\label{eq:a6}
-m_{\rm{eff}}\omega^{2}u_{n}=-k_{\rm{eff}}(2u_{n}-u_{n+1}-u_{n-1}),
\end{equation}
where
\begin{equation}
\begin{split}
\label{eq:a7}
m_{\rm{eff}}&=m_{1}-\frac{2k_{1}}{\omega^{2}}+\frac{4k^{2}_{1}}{(2k_{1}-m_{2}\omega^{2})\omega^{2}}-\frac{2k_{2}\cos^{2}\alpha}{\omega^{2}},\\
k_{\rm{eff}}&=\frac{k^{2}_{2}R^{2}\cos^{2}\alpha}{J\omega^{2}-2k_{2}R^{2}}.
\end{split}
\end{equation}
The $m_{\rm{eff}}$ and $k_{\rm{eff}}$ are the effective atomic mass and effective inter-atomic coupling of the system, which is equivalent to a 1D monatomic chain. As we can see that the effective coupling is only related to the rotational vibration resonance; the effective mass is not only  related to the rotational vibration resonance but also related to the translational vibration resonance of the outer-inter masses. Therefore, the dispersion relationship can be obtained by the effective parameters:
 \begin{equation}
\label{eq:a8}
\omega^{2}=4\frac{k_{\rm{eff}}}{m_{{\rm{eff}}}}\sin^{2}{\frac{qL}{2}}.
\end{equation}
Note since $k_{\rm{eff}}$ and $m_{{\rm{eff}}}$ are both functions of $\omega^2$, the band will split into three branches as we will see in the following.

\begin{figure}
\includegraphics[width=1\linewidth]{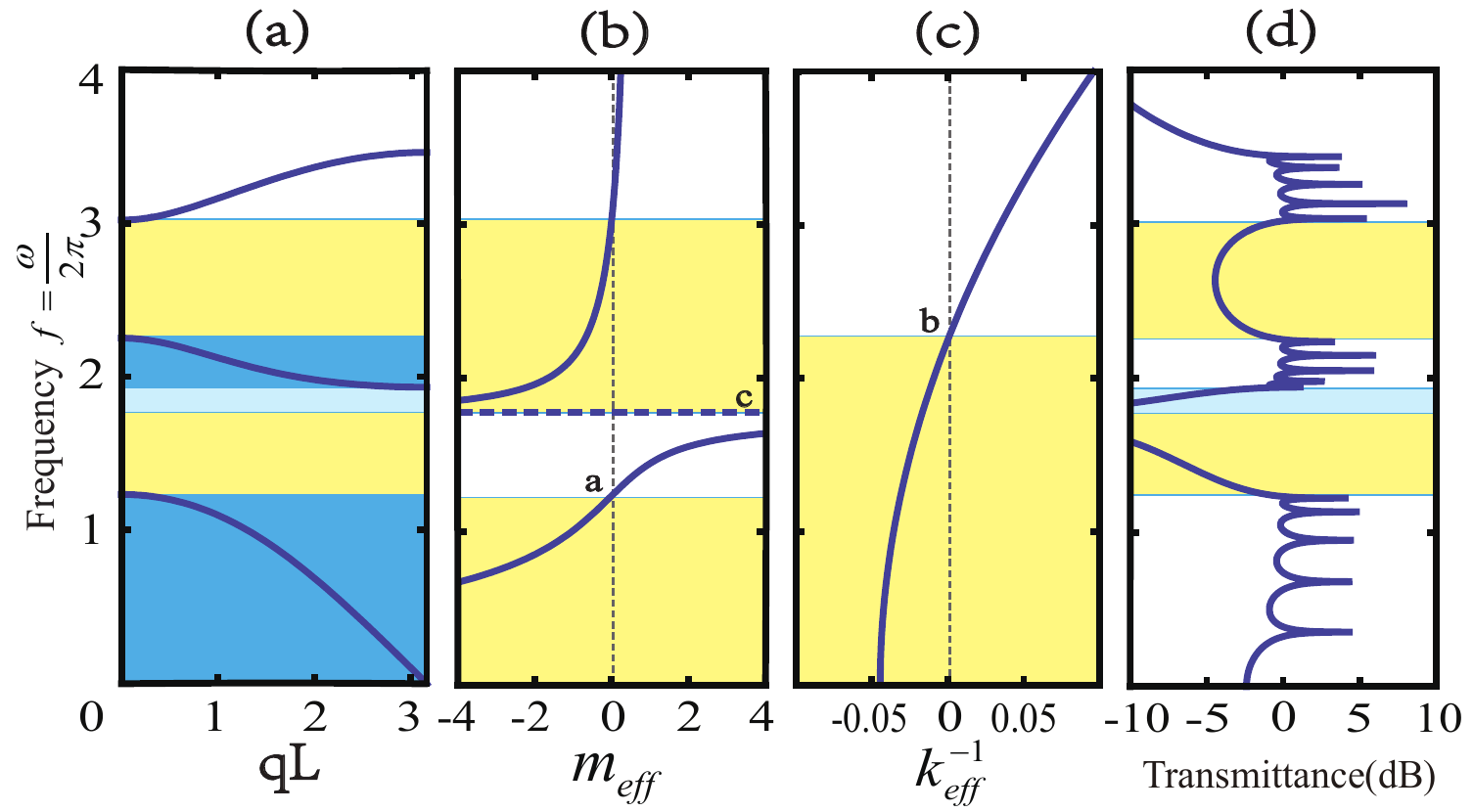}
\caption{\label{figure2} (a) The dispersion relation of the model (frequency $f$ = $\omega$/{2$\pi$}). (b) The effective mass $m_{\rm{eff}}$ $vs.$ frequency $f$. (c) The inverse of effective spring coupling $k^{-1}_{\rm{eff}}$ $vs.$ frequency $f$. (d) The transmittance spectrum of the system composed of five units (N=5). where $m_{1}=0.5$ $kg$, $m_{2}=0.5$ $kg$, $k_{1}=30$ $N/m$, $k_{2}=60$ $N/m$, $J=0.0015$ $kg\cdot m^2$, $R=0.05$ $m$, $\alpha=\pi/6$. Point a, b and c respectively represents the frequency when $m_{\rm{eff}}=0$, $k^{-1}_{\rm{eff}}=0$ and $m_{\rm{eff}}=\infty$. Yellow areas correspond to single negative area and blue areas correspond to double negative area.}
\end{figure}

Alternatively, we can also use the dynamic matrix of the model to obtain the dispersion relation. By letting the determinant of the system's dynamic matrix equal to zero:
\begin{align}
\text{Det}\left(\begin{matrix}\begin{smallmatrix}2k_{2}\cos^{2}\alpha+2k_{1}-m_{1}\omega^{2}&-2k_{1}
&k_{2}R\cos\alpha(1-e^{-iqL})\\
-2k_{1}&2k_{1}-m_{2}\omega^{2}&0\\
k_{2}R\cos\alpha(1-e^{iqL})&0&2k_{2}R^{2}-J\omega^{2}
\end{smallmatrix}\end{matrix}\right)=0,
\end{align}
we can get the dispersion relation $\omega(q)$ (frequency $f$=$\omega$/{2$\pi$}) of this model, as exemplified in Fig.~\ref{figure2}(a). Because each unit cell has three degrees of freedom,  there are three pass bands in the simplest Brillouin zone (the same result of the dispersion relation is obtained if using formula Eq.~(\ref{eq:a8})). We can see that the slope of the upper band is positive. So the direction of the group velocity and phase velocity are the same, which means that in this frequent range, the material has both positive mass and couplings, corresponding to the first quartile of Fig.~\ref{figure1}(b). However, the slope of the middle band and lower band are negative with positive phase velocity, which means that the group velocity is also negative and corresponds to the negative refraction in the third quartile of Fig.~\ref{figure1}(b).

But, for the two forbidden band gaps, we are not sure whether they are caused from translation resonance, rotation resonance or Bragg scattering, and which kinds of single negative are they. In order to further understand the underlying reason, we respectively check the effective mass $m_{\rm{eff}}$ and the inverse of effective coupling $k^{-1}_{\rm{eff}}$ vs. frequency $f$, as shown in Fig.~\ref{figure2}(b)(c) according to Eq.~(\ref{eq:a7}). The point $a$, $b$, $c$ means $m_{\rm{eff}}=0$, $k^{-1}_{\rm{eff}}=0$ and $m_{\rm{eff}}=\infty$, respectively. Yellow areas denote negative parameter regimes of effective mass and coupling. We can see that the middle and bottom bands correspond to both negative mass and negative spring coupling. That is to say, the two lower bands are double negative, agreeing well with our analysis above. It should be noticed that there appears a Bragg scattering induced gap below the bottom of the middle band, although within double negative regime. Besides, as displayed in Fig.~\ref{figure2}(b) and (b), the upper band gap and lower band gap corresponds to the single mass negative and single coupling negative, separately. Therefore the system will take on different properties under different frequency ranges.

Besides, we use the transfer matrix method to discuss the phononic transmission properties of the metamaterial. Our structure can be considered as an effective 1D monatomic lattice. Based on the previous calculation method~\cite{Yao2008Experimental}, the transmittance of the system is defined as $T=|\prod_{n=1}^N T_{n}|$, $T_{n}={u_{n}}/{u_{n-1}}$, 
$N$ is the number of unit cell. Thus, for the uniform 1D phononic metamaterial with $N$ unit, the recursive relation:
\begin{small}
\begin{equation}
\begin{split}
\label{eq:a9}
(2k_{\rm{eff}}-\omega^2m_{\rm{eff}})u_n&=k_{\rm{eff}}(u_{n+1}+u_{n-1}), n=1,2,...{N-1};\\
(k_{\rm{eff}}-\omega^2m_{\rm{eff}})u_N&=k_{\rm{eff}}u_{N-1},
\end{split}
\end{equation}
\end{small}
leads to the $n$-th transmittance $T_{n}$, expressed as:
\begin{equation}
\label{eq:a10}
T_{n}=\frac{k_{\rm{eff}}}{k_{\rm{eff}}(2-T_{n+1})-m_{\rm{eff}}{\omega^{2}}}, n=1,2...N
\end{equation}
with $T_{N+1}=1$.

We plot the transmittance $T=|\prod_{n=1}^N T_{n}|$ of the system (with 5 units, N=5) in Fig.~\ref{figure2}(d). 
From the colorized area of the figure, we can see that when the driving frequency matches well with the band gap, the system will have strong reflection and the transmittance is very small. The blue area is caused from Bragg scattering although it is in double negative range and yellow areas correspond to the band gap induced by the single negative properties which means that in the process of wave propagation, the wave present an evanescent wave of exponential decay leading to the vanishing transmittance within a certain frequency range.

In addition, as we can see from Fig.~\ref{figure2}, as the frequency changes, the structure takes on different properties. The frequency of point $a$, $b$ and $c$ in Fig.~\ref{figure2} corresponds to $m_{\rm{eff}}=0$, $k^{-1}_{\rm{eff}}=0$, $m_{\rm{eff}}=\infty$, respectively, which are at the bottom of the lower and higher frequency forbidden band gap and internal point of the lower frequency forbidden band gap. According to the formula of effective parameters Eq.~\ref{eq:a7}, the frequency of point $a$, $b$ and $c$ at long wavelength limit are :
\begin{small}
\begin{equation}
\begin{split}
\label{eq:a11}
f_a&=\frac{\omega_a}{2\pi} =\frac{1}{2\pi\sqrt{m_1m_2}}\bigg(k_1(m_1+m_2)+k_2m_2\cos^{2}\alpha\\
&+\sqrt{(k_1(m_1+m_2)+k_2m_2\cos^{2}\alpha)^{2}-4m_1m_2k_1k_2\cos^{2}\alpha}\bigg)^{\frac{1}{2}},
\end{split}
\end{equation}
\end{small}
\begin{equation}
\label{eq:a12}
f_b=\frac{\omega_b}{2\pi} =\frac{1}{\pi}\sqrt{\frac{R^{2}k_2}{2J}},
\end{equation}
\begin{equation}
\label{eq:a13}
f_c=\frac{\omega_c}{2\pi} =\frac{1}{\pi}\sqrt{\frac{k_1}{2m_2}}.
\end{equation}

\begin{figure}
\includegraphics[width=1\linewidth]{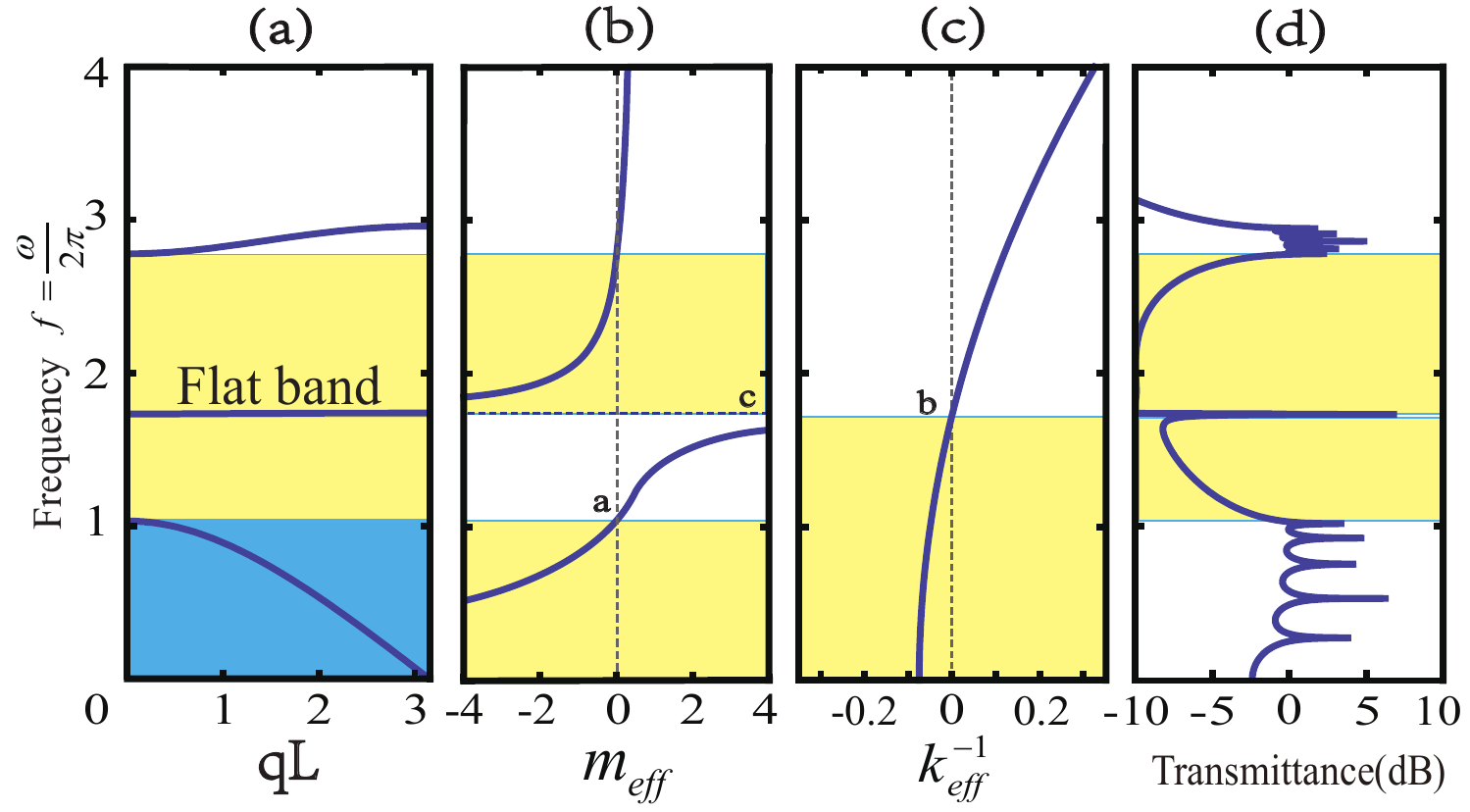}
\caption{\label{figure3} (a)The dispersion relation of the model with a flat band at the condition $f_c$ = $f_b$. (b) (c) (d) The effective mass, inverse of effective spring coupling and the transmittance spectrum for the system of five units (N=5) with lattice parameters $k_{2}=36$ $N/m$. Other parameters are the same as used in Fig.~\ref{figure2}.}
\end{figure}

When the resonant frequencies for translation and rotation are coincident ($f_b=f_c$) (see Fig.~\ref{figure3}), we have $m^{-1}_{\rm{eff}}=k^{-1}_{\rm{eff}}=0$. At this condition, the middle band become a flat band and the density of the state is very large. For the system with five units, at the frequency of flat band, there are five eigenvalues which means that the system has five kinds different mode of vibration. For all of the five kinds mode of vibration, we find that only $m_2$ and $\theta$ can move but $m_1$ stay still which indicates that the flat band is in accordance with that caused from dark state. In addition, the transmittance only has a single resonance peak at this resonance frequency. Higher or lower than this frequency, the system presents respectively single negative mass and single negative coupling. The flat band join these two single negative metamaterial regimes, so that the two phononic gaps merge into a single wide forbidden gap. Thus, tuning the effective negative parameters can cause total reflection effect and control wave transmission characteristics, which could realize the anti-vibration and the acoustic cloaking~\cite{Chen2007Acoustic,Cai2007Analysis,Cummer2008Scattering}.

We should point out that the structure proposed here could be used in a high frequency regime. For a typical nanostructrure with each unit cell containing up to hundreds of atoms, $m\sim10^{-26}-10^{-24}$ $kg$, which yields the frequency unit $[(k/m)^{1/2}]\sim 10^{12}-10^{13}$ $s^{-1}$. But if we choose the mass about $m\sim1$ $kg$, the frequency range will become around $1$ Hz, which can be steadily realized by a macroscopic vibrating experiment.

\section{Discusses}

\subsection{Zero-Index at Dirac point}

In the continuous limit (i.e., the long wave limit with infinitesimal $q$), $m_{\rm{eff}}{\partial^{2}u}/{\partial t^{2}}=-2k_{\rm{eff}}u(1-\cos{qL})$ can be expanded to be
\begin{equation}
\label{eq:a17}
\frac{m_{\rm{eff}}}{L}\frac{\partial^{2}u}{\partial t^{2}}=-k_{\rm{eff}}Lq^2u=k_{\rm{eff}}L\frac{\partial^{2}u}{\partial x^{2}},
\end{equation}
due to $q\rightarrow i\partial/\partial x$ for harmonic wave solution $u=\hat ue^{i(-qx+{\omega}t)}$. Thus, under this long wave limit, the vibration in 1D phononic metamaterials can be regarded as acoustic/elastic wave, and the discrete lattice can be regarded as a continuous medium. For such a medium, we know that the refractive index is defined as $n=v_s/v_p$ where $v_s$ is the reference sound velocity in air and $v_p=\omega/q=L\sqrt{k_{\rm{eff}}/m_{\rm{eff}}}$ is the phase velocity in the medium. Thus, the refractive index is
\begin{equation}
\label{eq:a14}
n\propto\sqrt{m_{\rm{eff}}/k_{\rm{eff}}}.
\end{equation}

In addition, the concept of impedance in wave physics is very important which can govern how a wave propagate in a system. In 1D acoustic metamaterials, the impedance $Z=\rho_{\rm{eff}} v_p$ with the line density $\rho_{\rm{eff}}=m_{\rm{eff}}/L$, so that
\begin{equation}
\label{eq:a15}
Z=\rho_{\rm{eff}} L\sqrt{k_{\rm{eff}}/m_{\rm{eff}}}=\sqrt{k_{\rm{eff}}m_{\rm{eff}}}.
\end{equation}

Derived from these formulas, if refractive index $n=0$ (the wave can propagate through the system without any changes), there are two cases to meet it. One is the single zero condition which means either one of the effective mass $m_{\rm{eff}}$ and the inverse of effective modulus $k^{-1}_{\rm{eff}}$ is zero, the other is double zero condition that is both of effective mass and inverse of effective spring coupling equal to zero. If the system is under the single zero condition, we can see that the wave impedance $Z=0$ or $\infty$, so that the wave can not pass into the system and propagate in it. But if it is under the double zero condition, the impedance $Z$ of the system can achieve a finite number so that the impedance of incident wave can matches well with the system and the wave will propagate through it, not cause a high reflection at the interface. At the same time, the refractive index $n$ is equal to zero. So a necessary for zero refractive index is the double zero condition.

\begin{figure}
\includegraphics[width=1\linewidth]{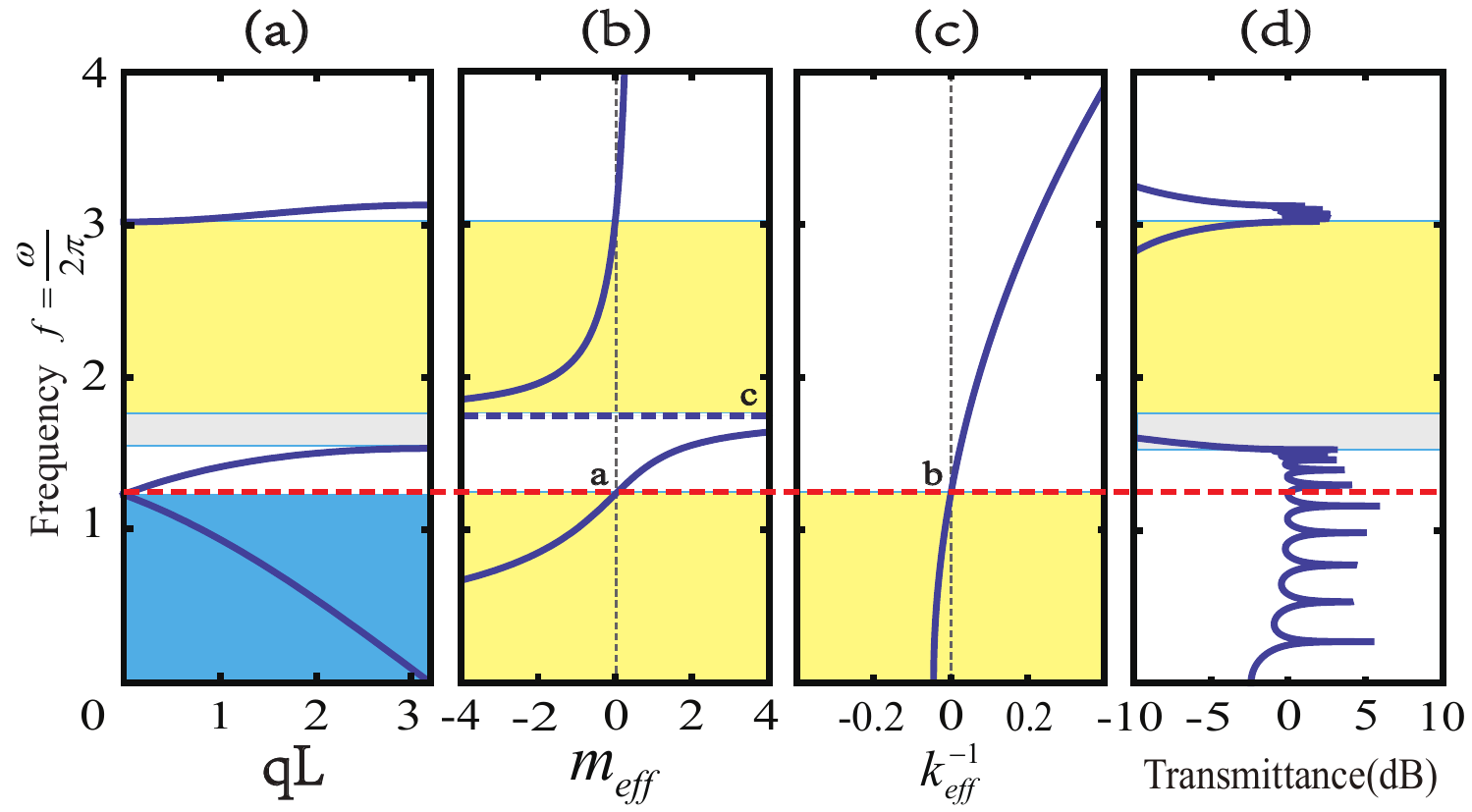}
\caption{\label{figure4} (a) The dispersion relation with a Dirac-like point $f_a=f_b$. (b) (c) (d) The effective mass, inverse of effective coupling and the transmittance spectrum of the system composed of five units (N=5) with lattice parameters $J=0.005$ $kg\cdot m^2$, the other parameters are the same as used in Fig.~\ref{figure2}. The red dash line indicates the zero value of $m_{\rm{eff}}$ and ${k^{-1}_{\rm{eff}}}$ at Dirac point. }
\end{figure}

For our system, if we want the refractive index $n=0$ which means the effective mass and the inverse of effective spring coupling are both zero at the same frequency, that is to say the frequency of point $a$ is the same as that of point $b$ ($f_a=f_b$). So we change the lattice parameters of the system and obtain zero refractive index in Fig.~\ref{figure4}(a). A Dirac-like point emerges at the critical point where the lower two pass bands degenerate. In order to further illustrate it is zero refractive index, we show the distribution of effective mass and the inverse of effective spring coupling in Fig.~\ref{figure4}(b) and Fig.~\ref{figure4}(c). Clearly, at the frequency of Dirac-like point, the value of $m_{\rm{eff}}$ and ${k^{-1}_{\rm{eff}}}$ are both zero. In addition, according to the Eq.~\ref{eq:a7}, using the lattice paraments of Fig.~\ref{figure4}, we calculate the effective mass and the inverse of effective coupling both equal to zero which meet the double zero condition at the frequency of Dirac-like point.

On the other hand, on the basis of Eq.~\ref{eq:a11} and Eq.~\ref{eq:a12}, we can obtain the formula of $m_1$ at Dirac-like point and then substitute it into Eq.~\ref{eq:a15}, the impedance become
\begin{equation}
\label{eq:a20}
Z=\frac{1}{\sqrt{2}R}\sqrt{Jk_2\cos^2{\alpha}(\frac{k_1k_2m^2_2}{(Jk_1/R^2-k_2m_2)^2}+\cos^2{\alpha})}.
\end{equation}
According to the lattice parameters of Fig.~\ref{figure4}, the impedance is calculated and equal to $7.5$, which is a finite number indicating that the wave can propagate into the system.

Besides, we also plot the transmittance spectrum of the system at Fig.~\ref{figure4}(d). But the Dirac-like point is not at the peak of the transmittance. The reason is that the boundary condition of the system we use to calculate the dispersion relation and transmittance is different. For the former we adopted periodic system but for the later we use a finite system. So there is a finite shift at the peak of the transmittance. As we increase the number of units, the peak of transmittance converges to that of the Dirac point.

\subsection{Topology}
\begin{figure}
\includegraphics[width=1\linewidth]{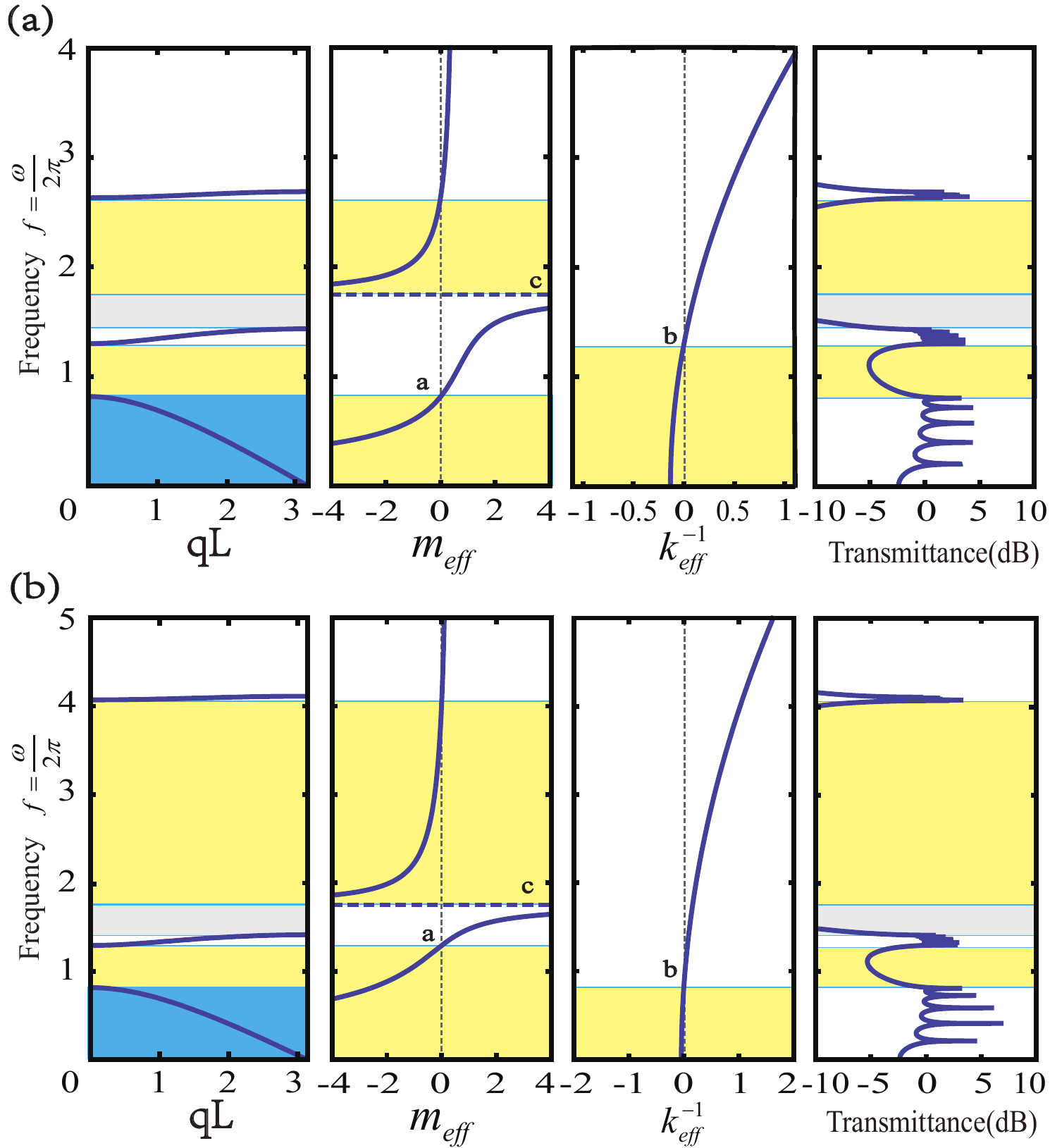}
\caption{\label{figure5} The dispersion relation, effective mass, inverse of effective coupling and the transmittance spectrum of the system composed of five units (N=5) with lattice parameters (a) $m_{1}=0.5$ $kg$, $k_{2}=20$ $N/m$, $J=0.0015$ $kg\cdot m^2$, (b) $m_{1}=0.25$ $kg$, $k_{2}=60$ $N/m$, $J=0.0114$ $kg\cdot m^2$, other parameters are the same as those in Fig.~\ref{figure2}.}
\end{figure}
In the previous section, we obtain the zero refractive index at Dirac-like point through $f_a=f_b$. Then we adjust the lattice parameters to two different setting: $f_a>f_b$, $f_a<f_b$ and find that the second band can flip and become double positive in both setting as shown in Fig.~\ref{figure5}. In Fig.~\ref{figure5}(a), the frequency of point $a$ is higher than the frequency of point $b$ ($f_a>f_b$). The gray gap area is caused by Bragg scattering. The lower frequency band gap at this case is due to the effective negative spring coupling. By further changing the parameters, we obtain Fig.~\ref{figure5}(b). The frequency of point $a$ is lower than the frequency of point $b$ ($f_a<f_b$). It is interesting to find that the lower band gap in this case is caused by negative effective mass within the same frequency range as in Fig.~\ref{figure5}(a). Therefore the two kinds of single negative band gap of Fig.~\ref{figure5}(a) and Fig.~\ref{figure5}(b) belong to different topological phases, respectively. The topological phase transition can take place when the system converts from effective elasticity negative (ENG) to effective mass negative (MNG), which is similar to the topological change between ENG ($\epsilon$ negative) and MNG ($\mu$ negative) in electromagnetism~\cite{Tan2014Photonic,Xue2016Topological,QSHE-TJ1,QSHE-TJ2}.

\begin{figure}
\includegraphics[width=1\linewidth]{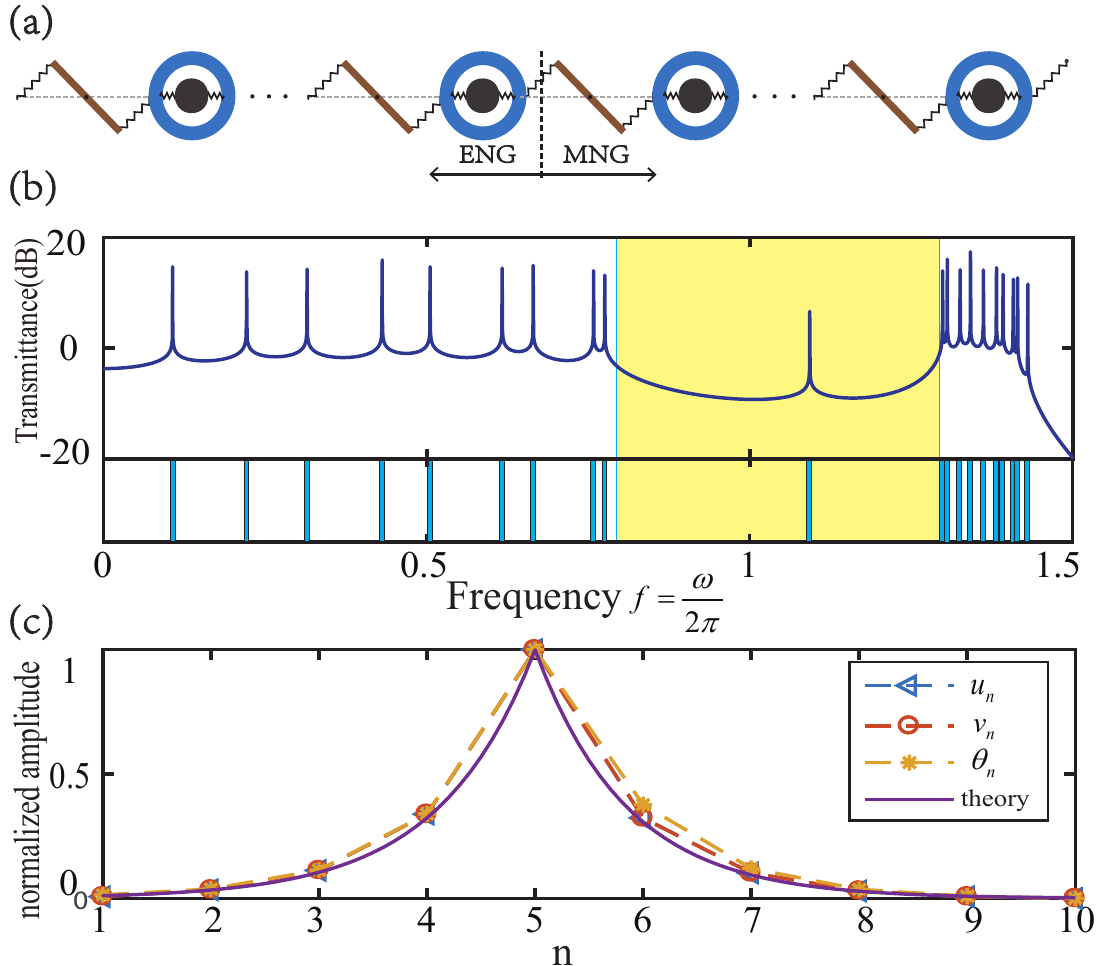}
\caption{\label{figure6} (a) The schematic spring-mass model of a 1D system to achieve the topological interface state with N=10. The left lattice is effective elasticity negative (ENG) and the right lattice is effective mass negative (MNG) in certain frequency range mentioned above. The last spring is added to change the edge-mode. (b) The transmittance spectrum and eigenvalue of the system composed of ten units (N=10). (c) The spatial profile of the eigenmodes of the interface state for the finite structure (N=10). The purple solid line is theoretical results of $u_n$. The blue, orange and yellow dashed line are respectively the eigenmode of $u_n$, $v_n$ and $\theta_n$. The parameters of the system are the same as those in Fig.~\ref{figure5}.}
\end{figure}

Topological interface states can exist in the interface between two simple lattices of distinct topological phases. In order to verify the topological phase transition between ENG and MNG phononic metamaterials, we construct a two-segment structure as shown in Fig.~\ref{figure6}(a) to manifest topological interface states in the lower band gap. The left five units are constructed as the system with effective negative spring coupling (ENG) in certain frequency range and the right five units are constructed as the system with effective negative mass (MNG).

In order to get a more intuitive view of the interface state, we plot the transmittance of the system in Fig.~\ref{figure6}(b). The upper part of the figure is the transmittance of the system and the lower part indicates the positions of eigenvalues. The eigenvalue matches well with the frequency of the resonance peaks. Clearly, there is a resonance peak within the gap, so-called mid-gap state. The spatial profile of the eigenmodes at the resonance frequency is shown in Fig.~\ref{figure6}(c). We can see that the resonant states at the five units and the amplitudes exponential decay on both sides of the interface between the ENG and MNG lattices. The resonance peak is reminiscent of the topological Jackiw-Rebbi interface state in the quantum field theory~\cite{TIbook}.

To understand the emergence of the interface state, we take the long wave limit as mentioned above. The 1D single lattice wave can be regarded as elastic wave and the wave number $q= \omega \sqrt{m_{\rm{eff}}/k_{\rm{eff}}}/L$. For a single negative system, $\sqrt{m_{\rm{eff}}/k_{\rm{eff}}}$ is an imaginary number. Thus, we can let $q\equiv -i\kappa$~\cite{Lee2012Acoustic}, where
\begin{equation}
\label{eq:a21}
\kappa=\omega \sqrt{|m_{\rm{eff}}/k_{\rm{eff}}|}/L.
\end{equation}
At the middle interface ($x=0$), the boundary condition is that the force and the displacement must be continuous, so that
\begin{equation}
\begin{split}
\label{eq:22}
&F_L=F_0e^{\kappa_L x+i\omega t},F_R=F_0e^{-\kappa_R x+i\omega t};\\
&u_L=u_0e^{\kappa_L x+i\omega t},u_R=u_0e^{-\kappa_R x+i\omega t}.
\end{split}
\end{equation}
According to the continuous version of Newton's equation of motion, $\rho_{\rm{eff}} d^2u/dt^2=-\partial F/\partial x$ [substituting $F=-k_{\rm{eff}}L\partial u/\partial x$ into Eq.~(\ref{eq:a17})], we obtain that:
$\omega^2m_{\rm{eff},L(R)}u_0/L=\pm \kappa_{L(R)} F_0$ should be hold on two sides of the interface.
 Therefore,
\begin{equation}
\label{eq:24}
\frac{m_{\rm{eff},L}}{\kappa_L}=-\frac{m_{\rm{eff}, R}}{\kappa_R} \Longrightarrow m_{\rm{eff},L}k_{\rm{eff},L}=m_{\rm{eff}, R}k_{\rm{eff},R}.
\end{equation}
Clearly, this condition indicates that the resonance peak meets the condition of impedance matching.
\begin{equation}
\label{eq:a16}
Z_{\emph{ENG}}=Z_{\emph{MNG}}.
\end{equation}
The closer the impedance of the two media is, the easier the wave will pass through the interface and cause a resonance peak. So, in our system, the wave impedance of the left lattice (ENG) and right lattice (MNG) are equal to each other at the resonance frequency.

For the system in Fig.~\ref{figure6}(a), the interface is at $n=5$. Using the lattice parameters of the system in Fig.~\ref{figure6}, we can calculate out the resonance frequency $f_{in}=\omega /(2\pi)$ satisfying Eq.~\ref{eq:24} as $f_{in}$=1.09 Hz. The penetration length of the interfacial evanescent wave is characterized by $1/\kappa$ from Eq.~(\ref{eq:a21}). The calculations from the continuous wave equation with impedance analysis agree well with the observation from the discretized lattice in Fig.~\ref{figure6}. At $f_{in}$, the left lattice and right lattice both have single negative property in the lower band gap, where their impedances match with each other to support the interface state.

\section{Conclusion}
In summary, we have shown that, the system we studied exhibits different effective properties, such as double positive, single negative or double negative parameters (mass and coupling), depending on the incident frequency. By calculating the transmission characteristics, we have observed that there is a low transmittance in the band gap which can be used to design vibration-absorptive material.
At the frequency where the effective mass and coupling are both infinite, a flat band emerges that will induce an extremely high density of states.
We have also achieved zero refraction index by adjusting the parameters for forming a Dirac-like point, where both effective mass $m_{\rm{eff}}=0$ and the inverse of effective spring coupling ${k^{-1}_{\rm{eff}}}=0$ such that the wave impedance is finite, which have wide application prospects on wave manipulations. Besides, the phenomenon of topological phase transition between negative mass and negative coupling has been studied in the low frequency band gap. Finally, we have analyzed the interface state arising from distinct topological phase between ENG and MNG.
These properties will deepen our understanding in both physics and application on the emerging concept of one dimensional topological phononic metamaterials.

\acknowledgements
This work is supported by the National Natural Science Foundation of China (No. 11775159), the Natural Science Foundation of Shanghai (No. 18ZR1442800), the National Youth 1000 Talents Program in China, and the Opening Project of Shanghai Key Laboratory of Special Artificial Microstructure Materials and Technology.

\bibliographystyle{apsrev} 
\bibliography{ref} 
\end{document}